\documentstyle[prl,aps,psfig,twocolumn]{revtex} 

\newcommand{\avg}[1]{\langle{#1}\rangle}
\newcommand{\td}[1]{{\tilde #1}}
\newcommand{\R}{{\cal R}}
\newcommand{\SL}{{\cal S}}
\newcommand{\W}{{\cal W}}
\newcommand{\re}{{\rm Re }}
\newcommand{\im}{{\rm Im }}

\begin{document}
\draft
\twocolumn[\hsize\textwidth\columnwidth\hsize\csname
@twocolumnfalse\endcsname
\title{Clustering of volatility as a multiscale phenomenon}
\author{Michele Pasquini and Maurizio Serva}
\address{Istituto Nazionale Fisica della Materia,
Universit\`a dell'Aquila, I-67010 Coppito, L'Aquila, Italy}
\address{Dipartimento di Matematica, Universit\`a 
dell'Aquila, I-67010 Coppito, L'Aquila, Italy}

\maketitle
\begin{abstract} 
The dynamics of prices in financial markets has been studied intensively
both experimentally (data analysis) and theoretically (models).
Nevertheless, a complete stochastic characterization of
volatility is still lacking.
What it is well known is that absolute returns 
have memory on a long time range,
this phenomenon is known as clustering of volatility.
In this paper we show that volatility correlations are power-laws with a
non-unique scaling exponent. 
This kind of multiscale phenomenology, which is well known to physicists since
it is relevant in fully developed turbulence 
and in disordered systems, is recently pointed out
for financial series.
Starting from historical returns series,
we have also derived the volatility distribution,
and the results are in agreement with a log-normal shape.
In our study we consider the New York Stock 
Exchange (NYSE) daily composite index 
closes (January 1966 to June 1998) and the 
US Dollar/Deutsch Mark (USD-DM) noon buying rates certified by the Federal
Reserve Bank of New York (October 1989 to September 1998).

\end{abstract} 
]
\narrowtext

\section{Introduction}

One of the most challenging problem in finance
is the stochastic characterization of market returns.
This topic not only has an academic relevance
but it also has an obvious technical interest. 
Think, for example, at the option pricing
models where distribution and correlations
of volatility play a central role. 

It is now well established that returns of the most important indices 
and foreign exchange markets
have a distribution with fat tails, and that they are
uncorrelated on lags larger than a single day,
in agreement with the hypothesis of efficient market.
On the contrary, the distribution of volatility 
and its correlations are still poorly understood. 
What it is known is that absolute returns 
(which are a measure of volatility) 
have memory on a long time range,
this phenomenon is known in financial literature as clustering of volatility.
Recent studies provide a strong evidence for power-law 
correlations for absolute returns \cite{Taylor,DGE,BB,CDL,Baillie,Pagan}.
Notice that in ARCH-GARCH approach \cite{Engle,Jorion,AB}
 volatility memory is longer than a single time step
but it decays exponentially, 
which implies that ARCH-GARCH modeling is inappropriate.
Indeed, GARCH models have been extended in order to take into account
this long memory properties \cite{DGE,Harvey,DLBC,BBM}.

In this paper we analyze the daily returns
of the the New York Stock Exchange (NYSE) composite index 
from January 1966 to June 1998, and the 
US Dollar/Deutsch Mark (USD-DM) noon buying rates certified by the Federal
Reserve Bank of New York from October 1989 to September 1998.
We not only find that volatility correlations are power-laws on
long time scales up to a year 
for NYSE index and six months for USD-DM exchange rate,
but, more important, that they exhibit
a non-unique exponent (multiscaling).
This kind of multiscale phenomenology is known to
be relevant in fully developed turbulence 
and in disordered systems \cite{PV}, and it is recently pointed out
for a financial series \cite{PSI,PSII}.
Our result is based on the fluctuation analysis of
a new class of variable that we call
{\it generalized cumulative absolute returns}.

The second main result of the paper is the study of
volatility probability distribution, which is derived
by means of Fourier transform analysis.
It is shown that it is well approximated by a log-normal distribution
for NYSE index, while a log-normal shape is a reasonable
fit only around the maximum for USD-DM rate.

The paper is organized as follows:
in section II we show that volatility 
has a long memory by considering
the autocorrelation
of absolute returns.
Nevertheless the power-law behavior cannot be inferred
by simply considering autocorrelations.
In order to have a sharper evidence for the nature
of the long memory phenomenon,
in section III, we perform a scaling analysis on the
standard deviation of a new class of observables, 
the generalized cumulative absolute returns.
This analysis implies power-law correlations
with non-unique exponent.
In section IV the attention is focused on volatility
probability distribution, computed from returns data 
by means of Fourier transform analysis,
which turns out to be log-normal at least for NYSE index.
In section V some final remarks can be found.

\section{Correlations for returns}
We consider the New York Stock Exchange (NYSE) daily composite index 
closes (January 1966 to June 1998) and the 
US Dollar/Deutsch Mark (USD-DM) noon buying rates certified by the Federal
Reserve Bank of New York (October 1989 to September 1998).
In the first case the dataset contains $8180$ quotes,
in the second $2264$.
The quantity we consider is the (de-trended) daily return, 
defined as 
\begin{equation}
r_t = \log{S_{t+1}\over{S_t}} - \avg{\log{S_{t+1}\over{S_t}}}
\end{equation}
where $S_t$ is the index quote or the exchange quote at time $t$.
The time $t$ ranges from 1 to $N$ where N is the total number of quotes
($8180$ for the NYSE index and $2264$ for the USD-DM exchange rate). 
The notation $\avg{\cdot}$ indicates the average over the whole sequence
of $N$ data.

As pointed out by several authors \cite{Clark,Mandelbrot,MS1},
the distribution of returns is leptokurtic.
In \cite{Mandelbrot}, it was firstly proposed a
symmetric L\'evy stable distribution and more recently 
in \cite{MS1} it is argued
that the distribution is L\'evy stable 
except for tails, which are approximately exponential.
The estimation is that the shape of a Gaussian 
is recovered only on longer scales, typically for monthly returns.

Let us introduce the autocorrelation for returns, defined as

\begin{equation}
C(L) = \avg{r_t r_{t+L}} -\avg{r_t} \avg{r_{t+L}} \ \ .
\label{autocorr}
\end{equation}

\begin{figure}
\begin{center}
\mbox{\psfig{file=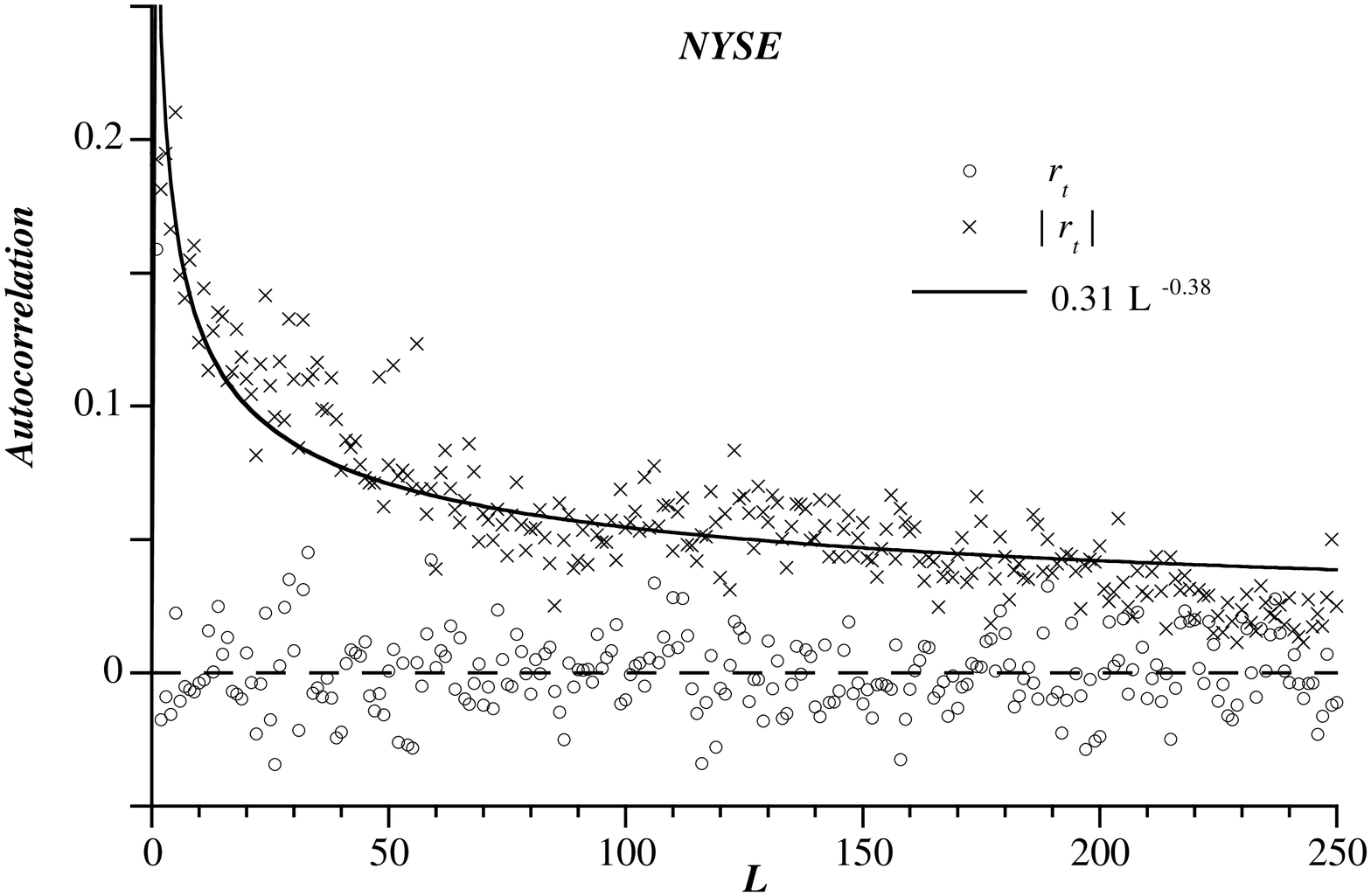,width=9cm}}
\mbox{\psfig{file=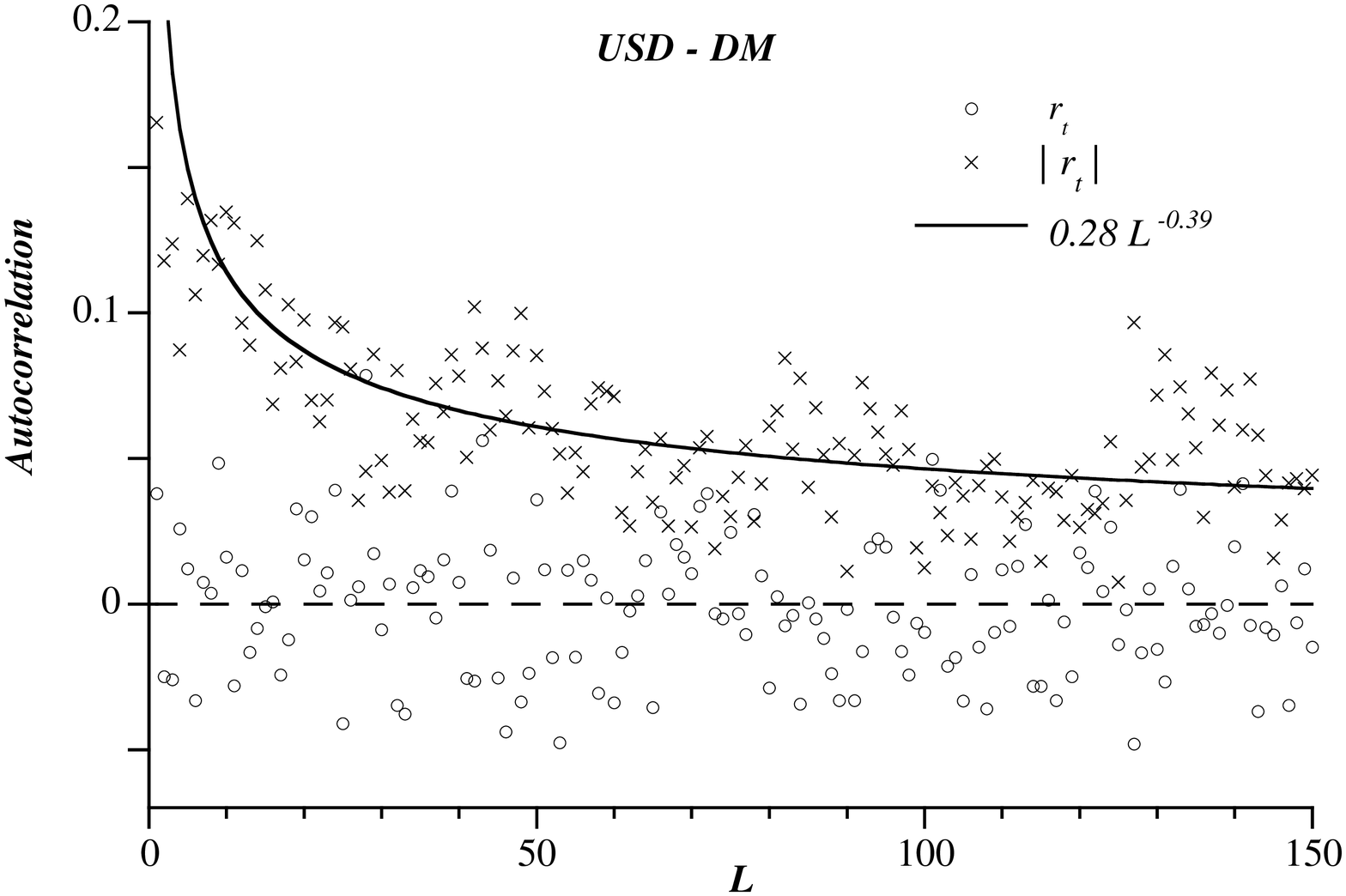,width=9cm}}
\end{center}
\caption{Autocorrelation $C(L,1)$ of $|r_t|$ (crosses) as a function
of the correlation length $L$, compared with the returns autocorrelation 
$C(L)$ (circles), for: a) NYSE index; b) USD-DM exchange rate.
The scale if fixed by autocorrelation equals to 1 at $L=0$.
The data for absolute returns are in agreement with a power-law
with exponent, respectively, $\alpha(1)\simeq 0.38$ for NYSE index and
$\alpha(1)\simeq 0.39$ for USD-DM rate, which are derived 
by the independent scaling analysis.}
\end{figure}

A direct numerical analysis of (\ref{autocorr})
for the NYSE index (fig. 1a) and for USD-DM rate (fig. 1b) 
shows that returns autocorrelation is a vanishing quantity for all $L$.
This simple evidence could induce to the wrong conclusion that
description is complete, i.e. returns are
i.i.d. variables whose distribution is a truncated Levy.
The situation is much more complicated, in fact,
even if returns autocorrelation vanishes,
one cannot conclude that returns are independent variables.
Independence implies that all functions of returns 
are uncorrelated variables.
This is known to be false, in fact volatility
have a long memory. On the other hand, 
the daily volatility is not directly observable, 
and informations about it can be derived
by means of absolute returns $\left|r_t\right|$.

It is useful to consider the following autocorrelation for 
powers of absolute returns
\begin{equation}
C(L,\gamma) = \avg{\left|r_t\right|^\gamma 
\left|r_{t+L}\right|^\gamma} 
-\avg{\left|r_t\right|^\gamma} \avg{\left|r_{t+L}\right|^\gamma}\ \ .
\label{clg}
\end{equation}
This quantity is plotted for $\gamma=1$ in fig. 1a (NYSE index)
and in fig. 1b (USD-DM exchange rate).
Unlike returns autocorrelation, it turns out to be a non vanishing 
quantity, at least up to  $L \simeq 150$
(see \cite{LCMPS,LGCMPS} and the references therein).
This is a clear evidence that it is not correct to assume
returns as independent random variables.

On the other hand, figs. 1 cannot give a satisfactory answer
about the shape of absolute returns autocorrelations.
In fact, data show a wide spread compatible with different
scaling hypothesis. In figs. 1 we have reported two 
power-law functions with exponents derived by scaling analysis,
which will be performed in the next section.
The proposed interpolations are consistent with numerical data.

\section{Scaling analysis}

In the previous section we have seen that,
consistently with the efficient market hypothesis, 
daily returns have no autocorrelations on lags 
larger than a single day.
This fact can be also checked by using of scaling analysis.
Consider the cumulative returns $\phi_t(L)$, defined as
the sum of $L$ successive returns $r_t, \dots, r_{t+L-1}$, 
divided by $L$ 

\begin{equation}
\phi_t(L) = {1\over L} \sum_{i=1}^L r_{t+i}
= {1\over L} \left[ \log{S_{t+L}\over{S_t}} 
- \avg{\log{S_{t+1}\over{S_t}}} \right]\ \ .
\end{equation}

One can define $N/L$ non overlapping variables of this type,
and compute the associated variance $Var \left( \phi(L) \right) $.
Assuming that $r_t$ are uncorrelated (or short range correlated),
it follows that $Var \left( \phi(L) \right) $ has a power-law behavior        
with exponent $\alpha=1$ for large $L$ (see Appendix A), i.e. 

\begin{equation}
Var \left(\phi(L) \right) \sim L^{-1}\ \ .
\end{equation}

The exponent $\alpha$ both for the NYSE index 
and USD-DM exchange market turns out to be around 1 
(see figs. 2 and also see \cite{MS2}), 
confirming that returns are uncorrelated.

On the contrary, this is not true for other quantities
related to absolute returns. In order to perform the
appropriate scaling analysis,
let us introduce the generalized cumulative absolute returns defined as
the sum of $L$ successive returns 
$\left|r_{t}\right|^\gamma, \dots, \left|r_{t+L-1}\right|^\gamma$, 
divided by $L$

\begin{equation}
\phi_t(L,\gamma) = {1\over L} \sum_{i=1}^L \left|r_{t+i}\right|^\gamma
\label{general}
\end{equation}
where $\gamma$ is a real exponent and, again,
these quantities are not overlapping.

\begin{figure}
\begin{center}
\mbox{\psfig{file=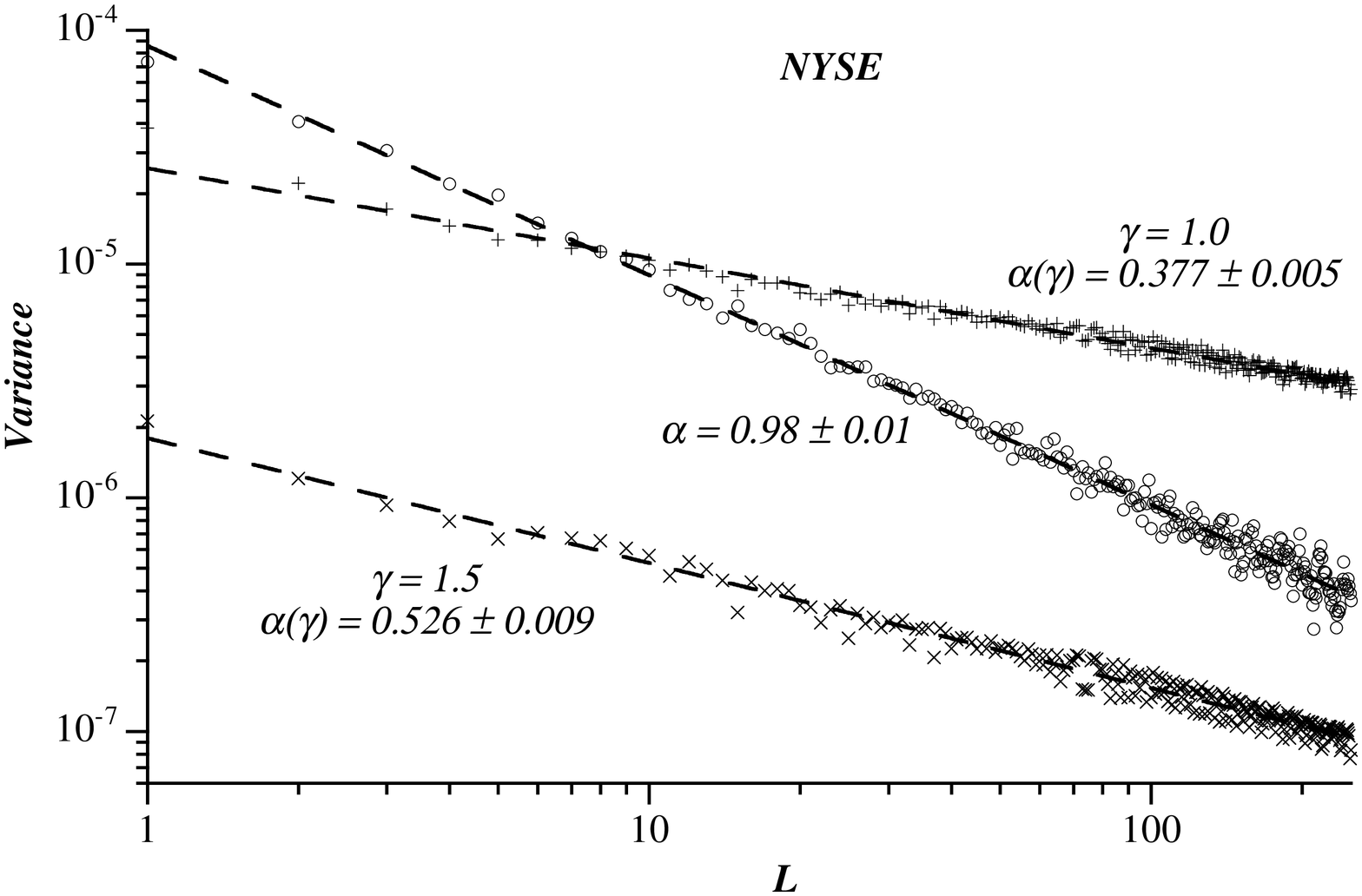,width=9cm}}
\mbox{\psfig{file=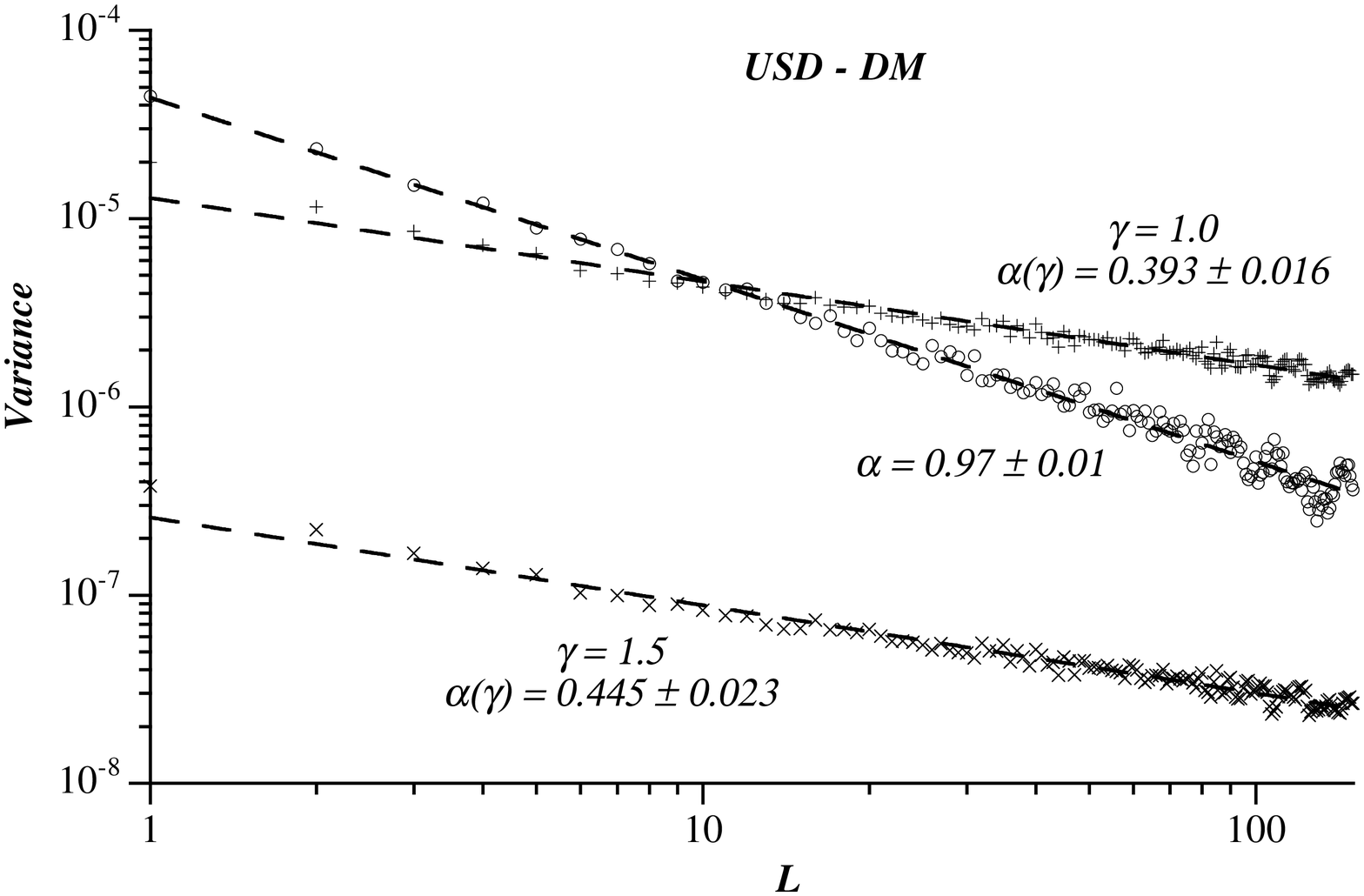,width=9cm}}
\end{center}
\caption{
Variance $Var \left( \phi(L,\gamma) \right) \sim L^{-\alpha(\gamma)}$ 
of the generalized
cumulative absolute returns as a function of $L$ on log-log scales
for $\gamma=1$ (crosses) and $\gamma=1.5$ (slanting crosses),
compared with the  variance 
$Var \left( \phi(L) \right) \sim L^{-\alpha}$ 
of the cumulative returns (circles),
for: a) NYSE index; b) USD-DM exchange rate.
The exponents of the best fit straight lines (dashed lines) are, 
respectively: 
$\alpha(1)=0.377\pm0.005$, $\alpha(1.5)=0.526\pm0.009$
and $\alpha=0.98\pm0.01$ for the NYSE index;
$\alpha(1)=0.393\pm0.016$, $\alpha(1.5)=0.445\pm0.023$
and $\alpha=0.97\pm0.01$ for the USD-DM exchange rate.
}
\end{figure}

\begin{figure}
\begin{center}
\mbox{\psfig{file=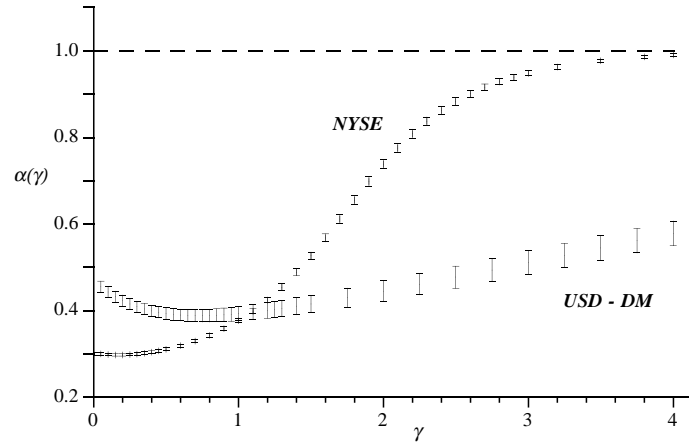,width=9cm}}
\end{center}
\caption{Scaling exponent $\alpha(\gamma)$ of the 
variance 
$Var \left( \phi(L,\gamma) \right) \sim L^{-\alpha(\gamma)}$ 
as a function of $\gamma$ for NYSE index and USD-DM rate,
where the bars represent the errors over the best fits.
An anomalous scaling ($\alpha(\gamma) < 1$) is shown 
for both cases.
}
\end{figure}

In appendix A we show that if the autocorrelation for 
powers of absolute returns (\ref{clg}) exhibits
a power-law with exponent $\alpha(\gamma)\le 1$ for large $L$, 
i.e. $C(L,\gamma) \sim L^{-\alpha(\gamma)}$, 
it would imply that 

\begin{equation}
Var \left(\phi(L,\gamma) \right) \sim L^{-\alpha(\gamma)} \ \ .
\label{varscal}
\end{equation}

On the contrary, if the $\left|r_t\right|^\gamma$
are short-range correlated or power-law correlated 
with an exponent $\alpha(\gamma)>1$,
we would not detect anomalous scaling in the analysis 
of variance, i.e. $Var \left(\phi(L,\gamma) \right) \sim L^{-1}$.

Our numerical analysis shows very sharply 
an anomalous power-law behaviour, after a very short transient time, 
in the range up to one year ($L=250$) for NYSE index (fig. 2a),
and up to six months ($L=150$) for the USD-DM 
exchange market (fig. 2b).
For larger $L$ the number of non overlapping 
variables $\phi(L,\gamma)$ becomes too small for a
statistical analysis, as revealed also by the increasing fluctuations
on variance $Var \left(\phi(L,\gamma) \right)$ as function of $L$.
The best fit straight lines are performed in the range, respectively,
$10 \le L \le 250$ for the NYSE index, and in the range
$10 \le L \le 150$ for the USD-DM rate.

The crucial result is that $\alpha(\gamma)$
is not a constant function of $\gamma$, 
showing the presence of different anomalous scales.
The interpretation is that different values of
$\gamma$ select different typical fluctuation sizes,
any of them being power-law correlated
with a different exponent.
The case $\gamma=0$ corresponds to
cumulative logarithm of absolute returns.
Approximately, in the region $\gamma \ge 4$ the averages 
are dominated by only few events, 
corresponding to very large returns and, therefore,
the statistics becomes insufficient.

In fig. 3, $\alpha(\gamma)$ is plotted as a function of $\gamma$
with error bars for both cases.
In the NYSE index case, the exponent $\alpha(\gamma)$
exhibits a large spread, reaching an ordinary 
scaling exponent
$\alpha(\gamma)=1$ for $\gamma \simeq 4$.
On the contrary, the USD-DM exponent turns out to be less
variable, rising slowly towards $\alpha(\gamma)=1$.

We would like to stress that the scaling analysis in figs. 2
definitively proves the power-law behaviour and precisely determine
the coefficients $\alpha(\gamma)$, 
while a direct analysis of the autocorrelations (as in figs. 1)
would not have provided an analogous
clear evidence for multiscale power-law behaviour, 
since the data show a wide spread 
compatible with different scaling hypothesis.

The anomalous power-law scaling can be eventually tested against 
the plot of autocorrelations. For instance, 
the autocorrelations of $r_t$ and of $|r_t|$ 
are plotted in figs. 1 as a function of the correlation length $L$,
and the full line, which is in a good agreement
with the data, is not a best fit
but it is a power-law whose exponent $\alpha(1)$ 
is obtained by the scaling analysis of the variance.

\section{Distribution of volatility}
All the discussion in previous section
concerns absolute returns.
An obvious question is: 'what is the relation with volatility?'.
The answer is not completely trivial, since from an operative
point of view, the volatility is often assumed to coincide
with the intra-day absolute cumulative return or, alternatively,
 with the implied volatility
which can be extracted from option prices.

Our point of view is that  the exact definition
of volatility cannot be independent from the
theoretical framework.
It is usually assumed that the volatility $\sigma_t$
is defined by

\begin{equation}
r_t = \ \sigma_t \ \omega_t
\label{rsw}
\end{equation}
where the $\omega_t$ are identically distributed
random variables with vanishing average and unitary variance.
The usual choice for the distribution of the $\omega_t$ 
is the normal Gaussian. This picture
is completed by assuming the probabilistic independence between
$\sigma_t $ and $\omega_t$.

In other terms, the returns series can be considered as a realization
of a random process based on a zero mean Gaussian, with a standard
deviation $\sigma_t$ that changes at each time step.
According to the above definition,
all the scaling property we have found on absolute returns
directly apply to volatility.

Volatility $\sigma_t$ is an {\it hidden} variable, since we can directly
evaluate only daily returns. Nevertheless, in appendix B
we show how to derive the volatility probability distribution $p(\sigma)$
starting from the returns series. The key point is to move the problem 
in the space of the characteristic functions (Fourier transforms).

The probability distribution $p(\sigma)$ is plotted in fig. 4,
both for the NYSE index and the USD-DM exchange rate.
The results corresponding to extreme values of volatility
($\sigma \simeq 0$ and $\sigma \simeq 0.02$) are not
confident due to insufficient statistics.

\begin{figure}
\begin{center}
\mbox{\psfig{file=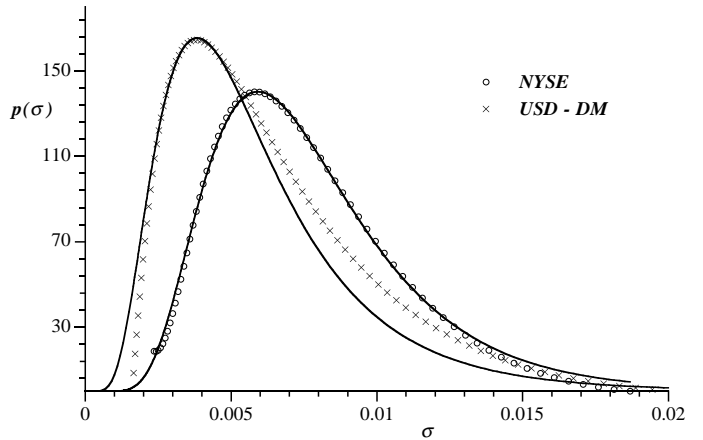,width=9cm}}
\end{center}
\caption{
Probability distribution $p(\sigma)$ of volatility for NYSE index
(circles) and USD-DM exchange rate (crosses),
fitted by log-normal distributions (\ref{lognorm})
with, respectively, $m=-4.94 \pm 0.01$ and $s=0.44 \pm 0.01$
for the NYSE index (fit performed in the range $0.0035 \le \sigma \le 0.01$),
and $m=-5.27 \pm 0.01$ and $s=0.54 \pm 0.01$
for the USD-DM rate (fit range $0.0025 \le \sigma \le 0.005$).
}
\end{figure}

The astonishing fact is 
that NYSE volatility distribution is well fitted
by a log-normal distribution \cite{LGCMPS,CLMPS}
\begin{equation}
p(\sigma) = {1\over {\sqrt{2 \pi} \ s \ \sigma}} \ \  
{\rm e}^{- {1\over 2} \left( {\log\sigma -m\over s} \right)^2}\ \ .
\label{lognorm}
\end{equation}
The fit is performed in the range $0.0035 \le \sigma \le 0.01$
and gives $m=-4.94 \pm 0.01$ and $s=0.44 \pm 0.01$, 
while the USD-DM volatility distribution
is consistent with a log-normal distribution 
with $m=-5.27 \pm 0.01$ and $s=0.54 \pm 0.01$
only in a narrow region around the maximum ($0.0025 \le \sigma \le 0.005$).

This unexpected log-normal shape for the volatility distribution suggests 
the existence of some underling multiplicative process
for volatility, at least for the NYSE index.
This result implies
that not only indices prices are multiplicative processes,
but also the associated returns.
On the other hand, the  USD-DM rate analysis might be affect
by insufficient statistics problems, which leads to an over-estimation
of distribution tail in the range $\sigma \simeq 0.01$.
Under this hypothesis, a log-normal shape could be 
consistent with the USD-DM volatility distribution,
and an underling multiplicative process might be present
also for foreign exchange returns.

A possible and reasonable tentative to explain
this peculiar behaviour for the volatility distribution
can be found in \cite{AMS}, where
a multiplicative cascade process for volatility is proposed,
borrowing well-known arguments from turbulence theory.

\section{Conclusions}
The first result we have found is that the scaling of variance
of the generalized cumulative absolute returns
is power-law with non-unique exponent,
for both the NYSE daily index and the USD-DM exchange rate.
This fact implies power-law correlations whose exponent
depends on the variable which is considered.
The main theoretical consequence is that models 
with exponential correlations, like ARCH-GARCH,
fails in describing
the dynamics of financial markets, and
that new models should  
account for the coexistence of
long memory with different scales.

The second result is that volatility distribution 
is log-normal, at least for NYSE index.
This fact suggests that volatility itself evolves as
a multiplicative process.

These two results show the existence of an
underling process that drives daily returns,
and indicates that new modelizations of financial markets 
have to look to returns as a subordinate process
of volatility.

\vskip 1cm
\noindent
{\bf Acknowledgements}

We thank Roberto Baviera, Rosario Mantegna and Angelo Vulpiani
for many interesting conversations concerning
data analysis and models for dynamics of prices.

\section*{Appendix A}

In this appendix we show that 
if the correlations $C(L,\gamma)$ exhibit a long range memory, 
$C(L,\gamma) \sim L^{-\alpha(\gamma)}$, then also
the variance $Var (\phi(L,\gamma))$ of the
{\it generalized cumulative absolute returns} 
behaves at large $L$ as $L^{-\alpha(\gamma)}$.

The explicit expression of variance is
$$
Var \left(\phi(L,\gamma)\right) =
  \frac{1}{L^2} \sum_{i=1}^L \sum_{j=1}^L
      \avg{|r_{t+i}|^\gamma |r_{t+j}|^\gamma} - 
      \avg{|r_{t+i}|^\gamma} \avg{|r_{t+j}|^\gamma} \,\,.
$$
Taking into account that $r_t$ is a stationary process, 
and using the definition of $C(L,\gamma)$ (\ref{clg}), one has:
$$
Var \left(\phi(L,\gamma)\right) =
{1\over L} C(0,\gamma)
+ {2\over L^2} \sum_{L \ge i>j \ge 1} C(i-j,\gamma) 
$$
where
$$ C(0,\gamma) = \avg{|r_t|^{2\gamma}} - \avg{|r_t|^\gamma}^2 \,\,. $$
The previous expression can be rewritten as
$$ 
Var \left(\phi(L,\gamma)\right) = {1\over L} C(0,\gamma)
+ {2\over L^2} \sum_{i=1}^{L-1} (L-i)\  C(i,\gamma) \,\,.
$$
Under the hypothesis $C(L,\gamma) \sim L^{-\alpha(\gamma)}$, 
one has for large $L$
$$ 
{2\over L^2} \sum_{i=1}^{L-1} (L-i) \ C(i,\gamma)
   \sim L^{-\alpha(\gamma)} 
$$
which leads to
$$ 
Var \left(\phi(L,\gamma)\right) = O(L^{-1}) + O(L^{-\alpha(\gamma)}) \,\,.
$$
For our data $\alpha(\gamma) \le 1$, and then
$$ 
Var \left(\phi(L,\gamma)\right) \sim L^{-\alpha(\gamma)} \,\,.
$$

On the contrary, if $\alpha(\gamma) > 1$ or worst, 
correlations exhibit a faster decay, the variance $Var (\phi(L,\gamma))$ 
would be a power-law with scaling exponent equals to $1$.

A similar sketch can be repeated for the cumulative returns
$\phi(L)$. In this case since correlation has a fast decay,
we have
$$ 
Var \left(\phi(L,\gamma)\right) \sim L^{-1} \,\,.
$$

\section*{Appendix B}

Let us introduce the variables $\R_t, \SL_t, \W_t$, defined as
$$
 \begin{array}{lll}
 \R_t = \log | r_t | \\
 \SL_t = \log \sigma_t \\
 \W_t = \log | \omega_t | 
\end{array}
$$
which are related among them by virtue of (\ref{rsw}) by
$$
\R_t = \SL_t + \W_t\ \ .
$$
For the  associated probability distributions 
(respectively $Q(\R), P(\SL), T(\W)$) the following relation holds
\begin{equation}
Q(\R) = \int_{-\infty}^{+\infty} d\SL \ P(\SL) T(\R-\SL)\ \ .
\label{integr}
\end{equation}
The distribution $P(\SL)$ retains full information on the volatility 
probability distribution $p(\sigma)$, 
since $p(\sigma)=P(\log\sigma)/\sigma$.

In order to derive from (\ref{integr}) 
an explicit expression for $P(\SL)$,
it is convenient to consider the characteristic functions 
(Fourier transforms) $\td{Q}(\td\R), \td{P}(\td\SL), \td{T}(\td\W)$ 
of $Q(\R), P(\SL), T(\W)$.
In fact, the following simple relation holds
$$
\td{Q}(\td\SL)= \td{P}(\td\SL) \td{T}(\td\SL)
$$
and the inverse Fourier transform gives
$$
P(\SL) = {1\over 2 \pi} \int_{-\infty}^{+\infty} d\td\SL \ 
{\td{Q}(\td\SL)\over \td{T}(\td\SL)} \ {\rm e}^{i \SL \td\SL}\ \ .
$$
Notice that $\td{Q}(\td\SL)$ and $\td{T}(\td\SL)$ are complex objects,
but we may consider only the real part of the integrand,
since the result of the integration has to be real
\begin{equation}
P(\SL) = {1\over 2 \pi} \int_{-\infty}^{+\infty} d\td\SL \ 
\re \left[
{\td{Q}(\td\SL)\over \td{T}(\td\SL)} \ {\rm e}^{ i \SL \td\SL}
\right] 
\label{integr2}
\end{equation}
where
$$
{\re \left[
{\td{Q}(\td\SL)\over \td{T}(\td\SL)} \ {\rm e}^{ i \SL \td\SL}
\right]} =
{\left( \re\td{Q} \ \re\td{T} +
\im\td{Q} \ \im\td{T} \right) \cos(\SL \td\SL)
\over {(\re\td{T})^2 + (\im\td{T})^2}}
+
$$
$$
 +
{ \left( \re\td{Q} \ \im\td{T} -
\im\td{Q} \ \re\td{T} \right) \sin(\SL \td\SL)
\over {(\re\td{T})^2 + (\im\td{T})^2}}\ \ .
$$

From a practical point of view, $\re\td{Q}(\td\SL)$ and $\im\td{Q}(\td\SL)$
can be directly computed from the returns series
$$
\re\td{Q}(\td\SL) = \int_{-\infty}^{+\infty} d\R \ Q(\R)\ \cos(\td\SL \R)
\ \simeq \ {1\over N} \sum_{t=1}^N \cos(\td\SL \R_t)
$$
$$
\im\td{Q}(\td\SL) = \int_{-\infty}^{+\infty} d\R \ Q(\R)\ \sin(\td\SL \R)
\ \simeq \ {1\over N} \sum_{t=1}^N \sin(\td\SL \R_t)\ \ .
$$

The Fourier transforms 
$\re\td{T}(\td\SL)$ and $\im\td{T}(\td\SL)$ can be evaluated numerically
starting from their definitions:
$$
\re\td{T}(\td\SL) = \int_{-\infty}^{+\infty} d\R \ T(\R)\ \cos(\td\SL \R)
$$
$$
\im\td{T}(\td\SL) = \int_{-\infty}^{+\infty} d\R \ T(\R)\ \sin(\td\SL \R)
$$
where
$$
T(\R) = \sqrt{2\over\pi} \ {\rm e}^{ \R - {1\over2}{\rm e}^{2 \R} }\ \ .
$$
Finally, the probability distribution $P(\SL)$, and then $p(\sigma)$,
can be computed via the numerical evaluation of integral (\ref{integr2}).

The key step of this procedure is the numerical inverse Fourier transform,
therefore the delicate point is the evaluation of the tails of the 
probability distribution $P(\SL)$, where the limited number of data
leads to spurious fluctuations.

\end{document}